\documentclass[flushrt,preprint]{aastex}
\usepackage{graphicx}
\usepackage{amsmath,amssymb}
\usepackage[hyphens]{url}
\usepackage{natbib}
\usepackage{lineno}
\begin{document}
%IJOC demands line numbers; uncomment next line
%\linenumbers
\sloppy
\title{Trends in U.~S. Storminess 1949--2009}
\shorttitle{U.~S. Storminess Trends 1949--2009}
\shortauthors{Canel and Katz}
\author{L. M. Canel}
\affil{Center for Urban Science + Progress}
\affil{New York University, Brooklyn, N. Y. 11201}
\author{J. I. Katz}
\affil{Department of Physics and McDonnell Center for the Space Sciences}
\affil{Washington University, St.~Louis, Mo. 63130}
\affil{Center for Urban Science + Progress}
\affil{New York University, Brooklyn, N. Y. 11201}
\affil{Tel.: 314-935-6202; Facs: 314-935-6219}
\email{katz@wuphys.wustl.edu}
\begin{abstract}
We use an extensive NOAA database of hourly precipitation data from 5995
stations in the 48 contiguous United States over the period 1949--2009 to
investigate possible trends in the frequency and severity of extreme
weather events, defined as periods of intense precipitation.  The frequency
and intensity of these events are quantified by a dimensionless storminess,
defined as the variance of the hourly rainfall at a site normalized by the
square of the mean rainfall at that site.  For 1722 stations with sufficient
data, we compute the rate of change of the logarithm of the storminess at
each station and set bounds on its mean (over stations) trend; use of the
logarithms weights trends at calm stations equally to those at stormy
stations and enhances the statistical power of the mean.  These results are
confirmed by reversing the order of averaging: first computing, for each
year, the mean (over stations) logarithm of the storminess, and then fitting
to its trend.  We set $2\sigma$ upper bounds of 0.001/y (doubling or halving
time scales $> 1000$ y) on any trend, increasing or decreasing.
\end{abstract}
\maketitle
\section{Introduction}
\label{sec:intro}
It is generally accepted that atmospheric and surface temperatures in most
or all climate zones, temporally averaged over shorter term and decadal
oscillations, have warmed since the 19th Century 
\citep{JM03,MW05,HRSL10,BEST}.  Although this issue has been settled to the
satisfaction of most climatologists, other questions remain open.  Many
studies \citep{G05,A06,P08,MZZH11,USKMO11,T11,G12,CR12,V13,W13,LCF15,S16}
and the Fifth IPCC consensus report \citep{IPCC1} have suggested that
warming has been or will be accompanied by an increased frequency of
``extreme weather events''.  These include periods of unusually high or low
precipitation, intense storms or droughts.  Although storms and droughts may
be considered opposites, one being defined or characterized by intense
precipitation and the other by the absence or deficit of precipitation, they
both correspond to increased variability of weather.  Statistically, they
both increase the variance and higher moments of the temporal distribution
of precipitation.

Counts and semiquantitative measures of severe storms such as hurricane
categories and the Fujita tornado scale have several drawbacks as tools
to measure trends in extreme weather events.  Detection of storms has
improved as technology has improved, biasing their statistics; an obvious
example is that satellite observations record hurricanes that never strike
land, whose detection in the pre-satellite era required their fortuitous
encounter with ships.  Quantification of a storm by sampled wind speeds is
only a rough approximation to the energy contained in its full
three-dimensional velocity field, and is biased by the development of
observing systems.  Perhaps most important, identified discrete storms are
few, limiting the statistical power of their data.

Severe storms are usually accompanied by short periods of intense rainfall.
Even tornadoes, not themselves sources of intense rain, are produced by
strong thunderstorms accompanied by rapid precipitation.  Precipitation data
include events over the entire spectrum of intensity, without arbitrary
thresholds (such as the distinction between a hurricane and a tropical
storm).  Most importantly, because precipitation data are available at a
large number of stations, at hourly frequency, over several decades, their
analysis may have great statistical power.

Previous studies of precipitation statistics 
\citep{G05,A06,MZZH11,USKMO11,BG11,K13a,K13b,FK14,A15,C15,LCF15,P15} have
mostly been concerned with the largest one-day or few-day (often five-day)
precipitation totals found in an annual or longer period.  Other studies
\citep{G12} have similarly studied the frequency of days with precipitation
over a high threshold or in a high range.  These data are valuable to civil
engineers and planners who must design storm water management systems, but
do not fully describe the statistics of precipitation.  Nor do they have the
statistical power and homogeneity of lengthy hourly precipitation records
that include the information contained in lesser events and dry periods.
The statistics of extreme events \citep{E48,J64,D83,K98,K99} do not take
advantage of the information present in lesser but more frequent events, nor
utilize all the data, and may depend on arbitrary choices of criteria and
thresholds.  Even studies of hourly data that deal only with their extremes
\citep{LM08,WML10,LMLO11} suffer from these disadvantages.

These factors also complicate combining information from multiple stations
that may be in different climatic regimes in which different definitions of
``extreme weather events'' might be appropriate; for example, thunderstorms
that occur nearly daily on the U.~S. Gulf Coast would be extraordinary in
the maritime climate of the Pacific Northwest.  Utilization of all available
data has been a powerful tool for extracting global warming from temperature
trends that may be inconsistent among stations that show stochastic local
and correlated regional variations \citep{IPCC1,BEST}.  Therefore, we define
metrics that combine information distributed throughout entire time series,
and that are weighted so that trends at calm and stormy stations contribute
comparably to the overall averages.  Our metrics utilize information from
periods of lower intensity precipitation as well as from intense storms
because trends in either may be manifestations of changing climate. 

For quantification of the ``storminess'' of climate, hourly data offer
information and insights lost when precipitation is averaged over 24 hour
periods.  A day of steady rain is not the same as a day during which an
intense storm produces the same amount of rain but concentrates it in one
or a few hours.  In order to take advantage of the full statistical power of
the extensive NOAA database containing (allowing for missing hours) about a
billion hourly data, this study uses information from more than a thousand
stations over the 61 years 1949--2009 to determine continent-wide trends.

The increase of water saturation vapor pressure with increasing temperature
(according ot the Clausius-Clapeyron equation) as the climate warms has been
predicted to increase the mean precipitation, the frequency of periods
(typically one or a few days) with precipitation above thresholds, and
record values of precipitation.  There is evidence of such effects
\citep{PAS07,BMH13,K13a,K13b,ALZLKB14,BHTPSC14,LCF15}.  However, we
distinguish trends in storminess (a measure of the concentration of
precipitation into extreme events) from trends in mean precipitation.
\section{Methods}
\label{sec:methods}
The NOAA database \citep{LCK} contains records of hourly precipitation at
5995 stations in the 48 contiguous United States from 1948 to 2009.  There
are few data from 1948, so we exclude that year.  A master (``COOP'') file
lists the intervals (typically of many years duration) during which each
station was nominally collecting data.  Individual station files contain
information for each day during which there was at least one hour with
measured precipitation and flag hours with missing data.  An individual
station file and the COOP file must be used together to determine the
hours when the station was actually collecting data.  We ignore
three stations for which no information is included in the COOP file.

Over the time period considered, almost all stations had at least some
months or years without data.  Indeed, according to the COOP file, only 1051
stations (of those for which there are hourly precipitation data) were
collecting data without extended interruption (these stations were still
subject to transient interruptions, typically of a few hours duration,
indicated in the individual station files but not in the COOP file).  Even
stations without extended interruptions generally do not have a complete
1949--2009 data series.  Many stations also moved small distances (typically
a few km) during the period analyzed.  We ignore these small moves.
%As an example, the first station in our list (COOP number 010008,
%in Abbeville, Alabama) moved 3.7 km in 1991, moved back close to its
%original position in 2001 and then moved another 0.6km in 2011.
%    [Distance calculations made using
%     http://www.movable-type.co.uk/scripts/latlong.html
%     and NOAA supplied latitudes and logitudes.]

%Finally, the data required additional light ``cleaning'':
We first addressed the problem of missing data.  The individual data files
have a code for isolated hours when the station was ``down'', and some files
used the same code for whole months when that station was not recording 
data.  Hours and days so indicated were removed from the data.  More
problematic were entries indicated as holding the cumulative precipitation
for an unspecified number of hours ending in the hour of the entry.  In such
cases the individual hourly values cannot be determined, so we ignored any
day with such an indicated hour.
%(RunParams.METHODNOD=2).

It is also necessary to ensure that accurate counts of hours of
data collection in order to compute valid averages.
The absence of an entry in a station data file may be due to either the
station being down or the absence of precipitation.
Unfortunately, the station data collection times obtained from the COOP
file are sometimes inconsistent with the precipitation
data in the individual station data files.  This can be in either direction:
a) precipitation data shown when the COOP file indicates that the station was
down that day, or b) no data in the station files for months and in places
where it is very implausible that there was no precipitation although the
COOP file indicates the station was up.  We
\begin{description}
\item[a)] ignore any months for which the COOP file indicates the station
was down during a period (within that month) but during which the data file
indicates data were collected;
%fact, data collection started but no rain fell so nothing was recorded in
%the station data file)
\item[b)] ignore data for complete years during which the individual
station file lists no precipitation %rain as falling during the calendar year
(this is very unlikely to be true for any location in the 48 states).  We
cannot distinguish \emph{individual months} for which the COOP file
indicates data were collected, but that might be missing data, from months
that actually had zero precipitation.  We do not exclude such months because
many stations, particularly in the Southwest and California, have months
without any precipitation, so that the complete absence of recorded
precipitation in a month is not an indication that a station was down.
\end{description}

A second set of cuts was made to reduce bias introduced by some types of
missing data:
\begin{description}
\item[a)] We ignore data from a station for a calendar year during which the
station was not up at least 80\% of the days; the absence of data for some
seasons would introduce bias.
%[Set in RunParams.MINNHRSUPTIMEPERYEARTOINCLUDE]
\item[b)] In order to study the variation of precipitation statistics as the
climate warms, we further restrict consideration to stations which have a
long duration of data collection.  We require that there be at least 30
%[set in RunParams.MINYEARSTOTALENOUGHDATA=15]
years of data spread over at least four %[Run.MINSPREADWINDOWENDSYRS ==33]
eleven year Solar cycles (to minimize any spurious trends resulting from
possible Solar cycle effects on weather),
%[RunParams.WINDOWLENGTHYRS]
with at least 6 %[RunParams.MINYEARSENOUGHDATAINWINDOW]
valid years (with the station up at least 80\% of the days) of data in each
cycle after earlier cuts have been made.
\end{description}

After subjecting the data in \cite{LCK} to these cuts, 1722 stations remain.
This gives our results statistical power not found in a preliminary study
\citep{MK13} of 13 stations that were considered individually.  That earlier
study found a nominally very significant ($5\sigma$) trend in storminess
at one station, a significance that depends on the assumption of a normal
distribution of annual storminess at each station about its mean trend.

The dimensionless normalized $n$-th moment for station $j$ is defined:
\begin{equation}
\label{norm}
M_{j,n,T} \equiv {\sum_{i \in T} (p_{j,i} - \langle p_{j,i} 
\rangle_{i \in T})^n \over N_T \langle p_{j,i} \rangle_{i \in J}^n},
\end{equation}
where the $p_{j,i}$ are the measured hourly data, including hours when there
was no precipitation, $i$ denotes the date and hour of measurement, $T$
denotes the temporal interval (the year) over which the moment is averaged,
$N_T$ is the number of valid data included in the sum in the numerator
(missing data are ignored), $J$ is the set of all valid data (over the
entire period 1949--2009), containing $N_J$ elements, for station $j$.
$N_T$ is generally less than the number of hours in $T$ and $N_J$ is less
than the number of hours in $J$ because some data are missing and some
months or years of data are rejected because of inconsistency with the COOP
file or to minimize bias from incompletely sampled years.  The mean
precipitation
\begin{equation}
\langle p_{j,i} \rangle_{i \in J} \equiv {1 \over N_J} \sum_{i \in J}
p_{j,i}.
\end{equation}

We define the ``storminess'' as the normalized second moment $M_{j,2,T}$ for
the station $j$ and year $T$.  Normalization distinguishes storminess from
wetness or dryness, and permits comparison of storminess at wet and dry
stations.  Wet stations may be either stormy (on the Gulf Coast) or calm (in
the Pacific Northwest), while dry stations are generally stormy (in the
desert Southwest) because what precipitation they do receive comes in
infrequent storms.
%where the second moment was first averaged over a
%year and then its annual averages were again averaged over a decade.

Normalization by the mean over the entire record minimizes any artefacts
resulting from shorter time scale variations of the mean precipitation.
Such phenomena may be aliased into long term trends if folded into bins such
as calendrical decades.  By fitting a linear trend to the entire record of
annual storminess at each station we minimize effects of phenomena on
decadal or shorter time scales.  It is still, with only 61 years of data,
not possible to distinguish the effects of phenomena on that, or longer,
time scales ({\it e.g.,\/} the Little Ice Age) from ``genuine'' secular
trends.  This is a well-known problem in all, necessarily finite,
geophysical time series \citep{MW69} because they are characterized by noise
with a ``pink'' spectrum.

In averaging over the 1722 surviving stations we have not attempted to allow
for the spatial density with which they sample climatic information, in
contrast to the work of \cite{BEST} with temperature data.  We question the
utility of doing so because the spatial scale on which climate varies is very 
nonuniform.  Large areas (such as the U.~S. Midwest) may have similar
climate, but in mountainous areas and near coasts, especially the U.~S.
Pacific Coast, sites separated by one km may have different, and perhaps
independently varying, climate, as is familiar to residents of San
Francisco.  It is not straightforward to define an unbiased algorithm to
account for this.  In our
earlier study of drought \citep{FCKK16} the site-averaged and area-averaged
results agreed to about 10\% despite a non-uniform distribution of stations.
\section{Results and Discussion}

\begin{figure}[h!]
\centering
\includegraphics[width=3.8in,angle=-90]{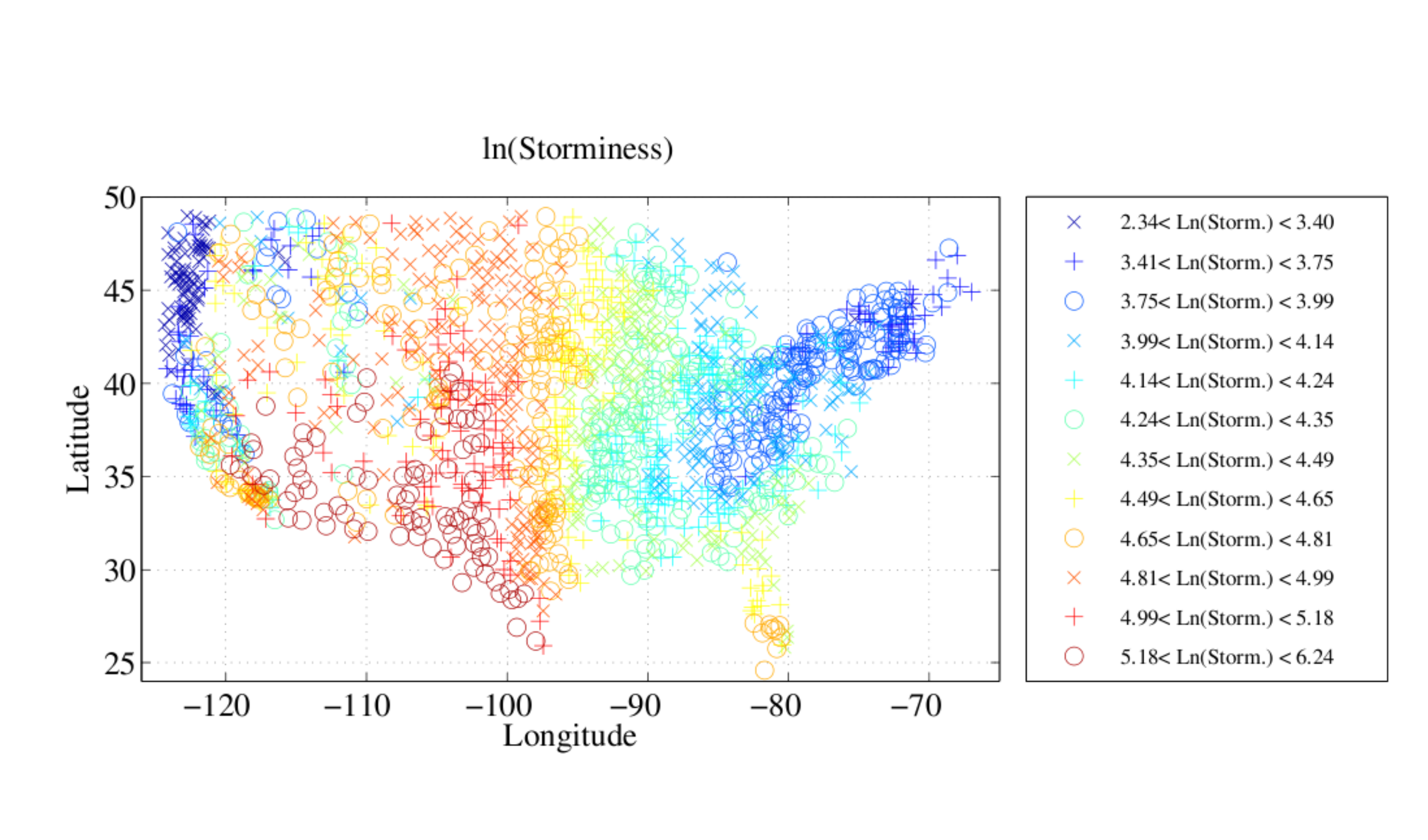}
\caption{\label{storminessmap}Natural logarithms of mean (1949--2009)
storminess at 1722 stations.  Symbols indicate the percentile ranks of the
storminess, divided into vigintiles (0--5\%, 5--10\%, 90--95\%, 95--100\%)
and deciles (10--20\%,$\ldots$,80--90\%).}
\end{figure}

Fig.~\ref{storminessmap} shows the mean values of the natural logarithms of
the storminess at our 1722 stations.  Storminess is high in regions (the 
Southwestern desert, Californian coast and western Great Plains) where
precipitation is concentrated both seasonally and into intense storms.
Storminess is low in regions, such as the Appalachian Mountains and the
Northeast, where precipitation is frequent, and especially in the maritime
climate of the Pacific Northwest where precipitation occurs as omnipresent
drizzle.  There is also a strong east-west gradient of storminess in the
Great Plains, dividing humid regions from drier but occasionally stormier
regions requiring dry farming.

For each station we make a linear fit to the logarithm of the storminess as
a function of time; the slope is the time derivative of $\ln{(\mathrm
{storminess})}$.  Logarithms are used, in part, because we are interested
in the percentage (or fractional) rate of change in storminess.  This
permits comparison of trends at stations whose storminesses may differ by
large factors (as much as 50); a significant trend would be scientifically
interesting (and might be practically important) whether it occurred at a
stormy or a calm station.

\begin{figure}[h!]
\centering
\includegraphics[width=3.7in,angle=-90]{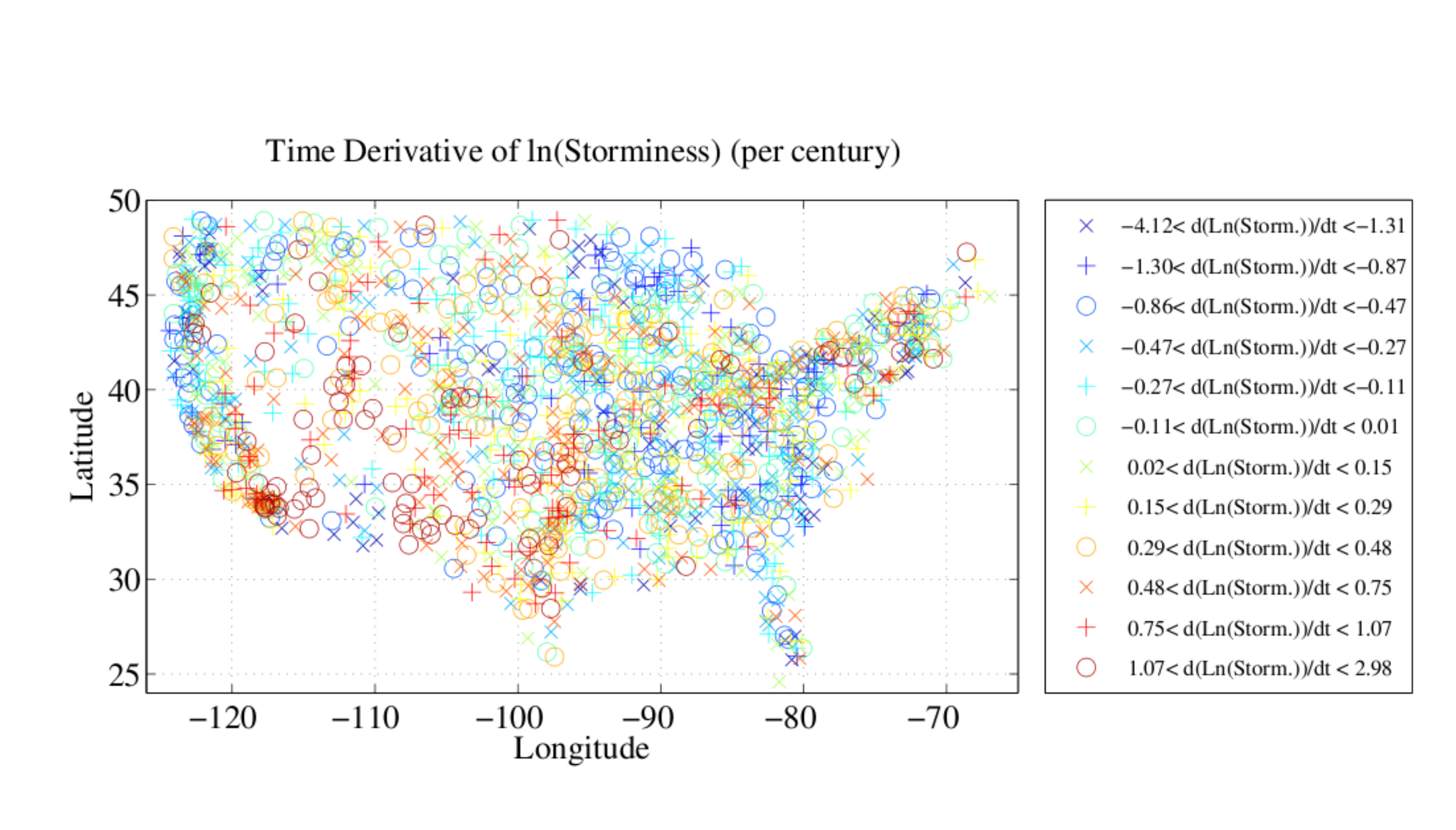}
\caption{\label{slopemap}Rates of change, per century, of the natural
logarithms of the storminess at 1722 stations.  Symbols indicate the
percentile ranks of the slopes, divided into vigintiles (0--5\%, 5--10\%,
90--95\%, 95--100\%) and deciles (10--20\%,$\ldots$,80--90\%).  There is
much more spatial heterogeneity than for the mean storminess
(Fig.~\ref{storminessmap}), with less correlation among neighboring stations.
There are hints of decreasing
trends in the Pacific Northwest and around Lake Superior and of increasing
trends in the Intermountain Basin and the southern Great Plains.  More
rapidly increasing storminess is found in the Los Angeles area.}
\end{figure}

Logarithms are used also because their mean slope is not dominated by those
of a few very stormy stations; a 10\% change (for example) at a calm station
contributes as much to the mean slope (and to our understanding of climate
change) as a 10\% change at a stormy station.  The uncertainty in the mean
slope is about $1/\sqrt{1722}$ of the standard deviation of the individual
(logarithmic) slopes, not $1/\sqrt{N_{stormy}}$ where $N_{stormy}$, a few
dozen, is the number of stormy stations that would contribute most of the
variance in the slopes of the raw storminess.  The results are shown in
Fig.~\ref{slopemap}.

\begin{figure}[h!]
\centering
\includegraphics[width=6in]{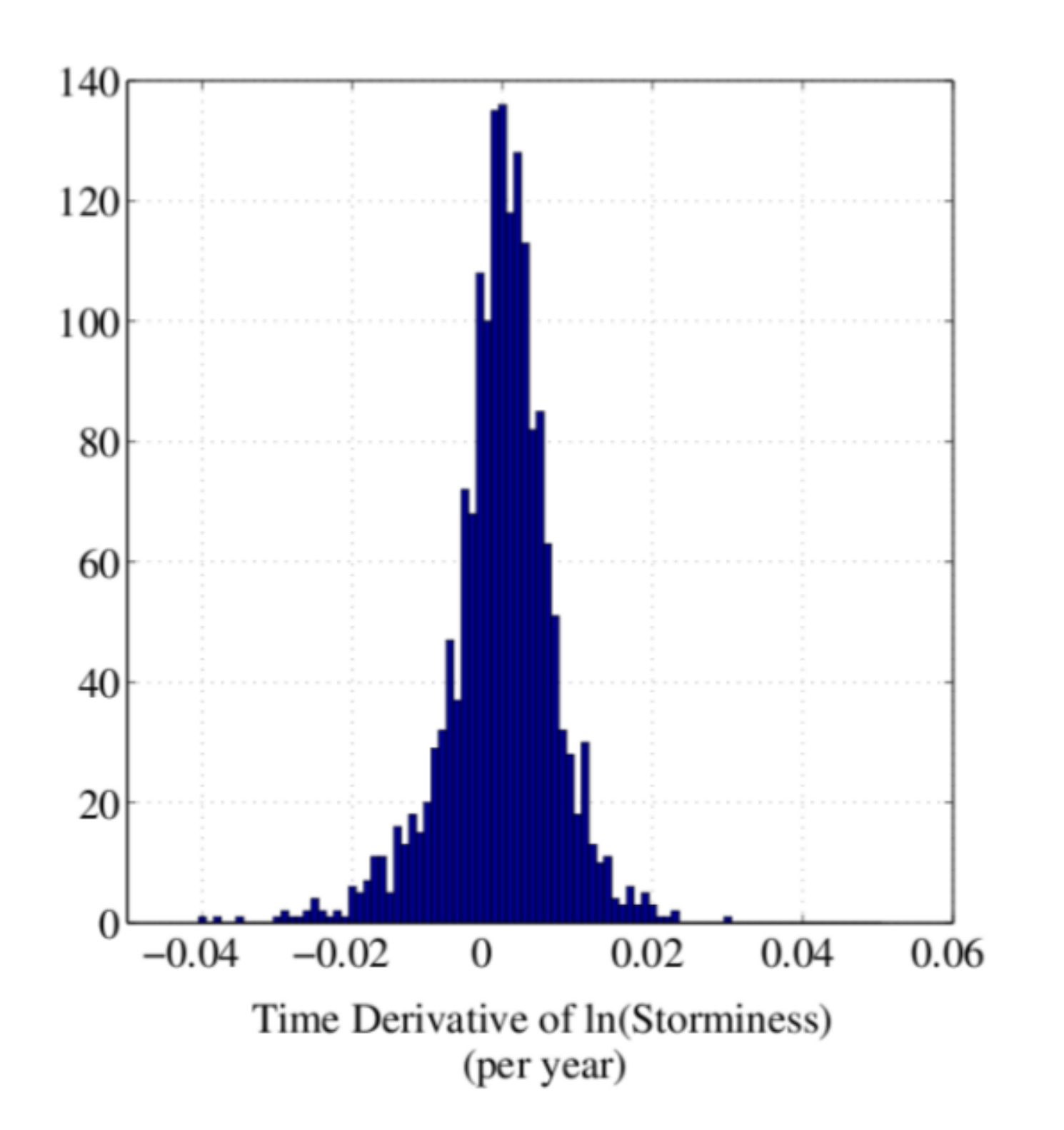}
\caption{\label{histogram}Distribution of slopes of logarithms of storminess
at 1722 stations in the 48 contiguous United States.  The mean is $(-2.8 \pm
1.7) \times 10^{-4}$/y, consistent with zero, and indicating the storminess
in the period 1949--2009 for which we have data is varying on an $e$-folding
time scale of no less than $\sim 1000$ years.  The skewness of the
distribution is $-0.67 \pm 0.06$ and its excess (compared to a Gaussian)
kurtosis is $2.79 \pm 0.12$, both of which are, at least nominally, very
statistically significant.  They suggest strong local or regional trends,
which are consistent with the existence of a variety of correlated (in space
and time) weather variations on all temporal and spatial scales.}
\end{figure}

Fig.~\ref{histogram} shows a histogram of the distribution of slopes (time
derivatives) of the logarithms of the storminess.  The mean slope is $(-2.8
\pm 1.7) \times 10^{-4}$/y, where the nominal standard deviation assumes
independent random variables, and the median slope is $1.5 \times
10^{-4}$/y.  The mean slope nominally differs from zero by $-1.6 \sigma$,
which is not significant.  The nominal $2\sigma$ range is $(-6.2,0.6) \times
10^{-4}$/y, corresponding to a nominal range from exponential decrease of
storminess with a halving time of 1000 y to exponential increase of
storminess with a doubling time of 10,000 y (this extremely long $2\sigma$
bound is an artefact of the fact that the mean is nearly $-2\sigma$; the
$3\sigma$ bound is about 3000 y).  Qualitatively, this may be described by
saying that the characteristic ($e$-folding) time of variation of the mean
(across the 48 states) storminess is no less than a millenium.  It is not
possible to exclude more rapid local trends.

Most of the logarithmic slopes are small, with characteristic time scales of
variation (defined as the reciprocal of the slope of the logarithm) of
centuries.  Negative and positive slopes are nearly balanced in number and
magnitude, and the mean slope is much less than the magnitudes of the slopes
at the stations with the most rapid change.  Although it cannot be proven,
the large differences among neighboring sites, even in the same climate
regimes, are consistent with the hypothesis that the observed slopes are
mostly the consequence of stochastic local events rather than systematic
trends.  However, by averaging over a large number of stations we can set
tight bounds on the mean trend.  In the period 1949--2009, during which
global warming was comparatively rapid, the \emph{mean} storminess in the 48
contiguous states changed comparatively slowly, if at all.

\begin{figure}[h!]
\centering
\includegraphics[width=6in]{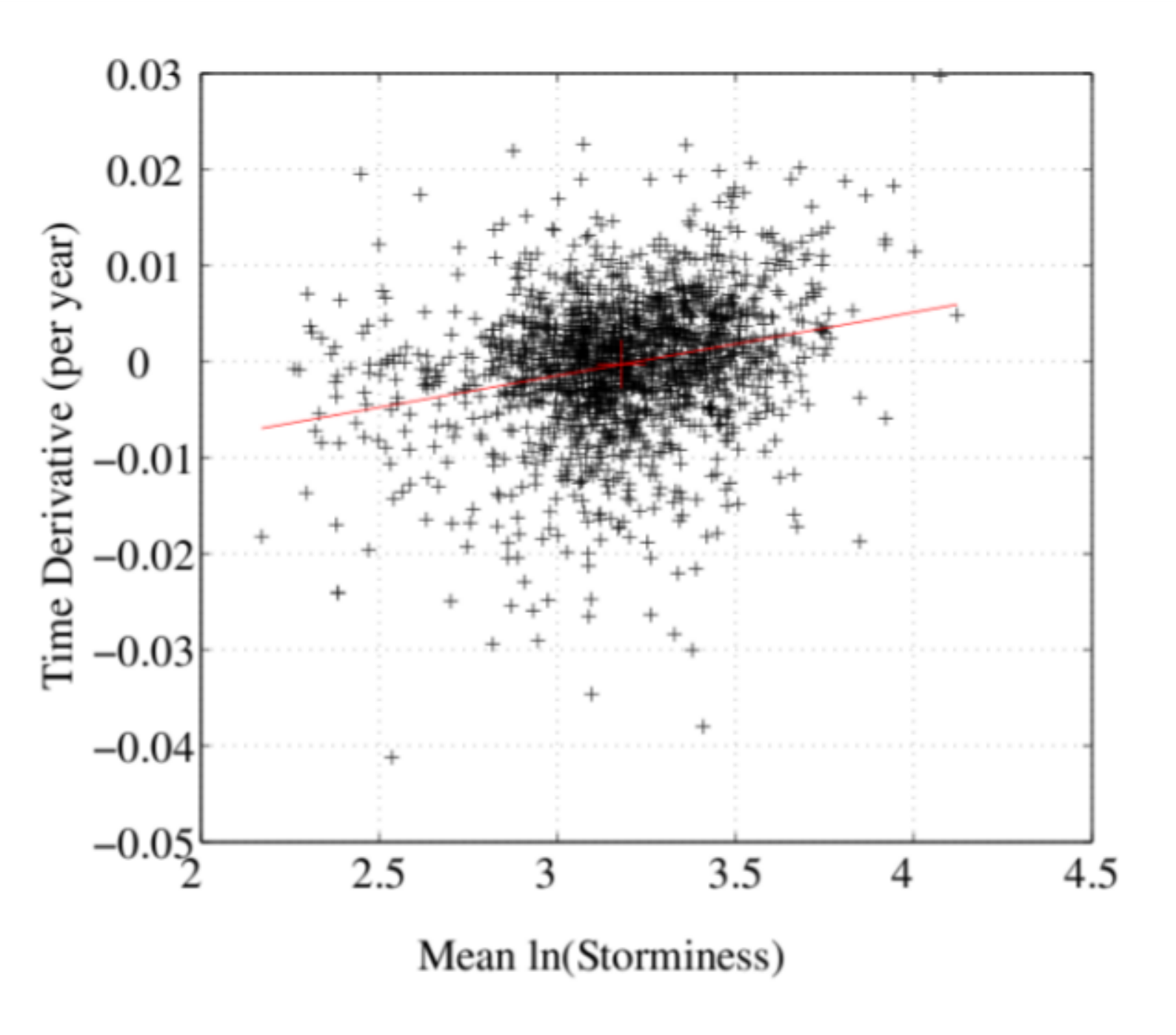}
\caption{\label{scatterplot}Logarithmic slopes of storminess {\it vs.\/}
mean storminess at 1722 stations in the 48 contiguous United States.  The
red line is a linear fit.  The Pearson's correlation coefficient is 0.25;
the correlation only accounts for a small fraction of the station-to-station
variation in slopes.}
\end{figure}

Fig.~\ref{scatterplot} plots the logarithmic slope of the storminess at
each station against its mean value.  There is a weak correlation (Pearson's
correlation coefficient is 0.25) that is nominally significant, but the
random scatter is greater than any trend related to the mean storminess, and
the fitted trend only accounts for a small part of the scatter.

\begin{figure}[h!]
\centering
\includegraphics[width=4in,angle=-90]{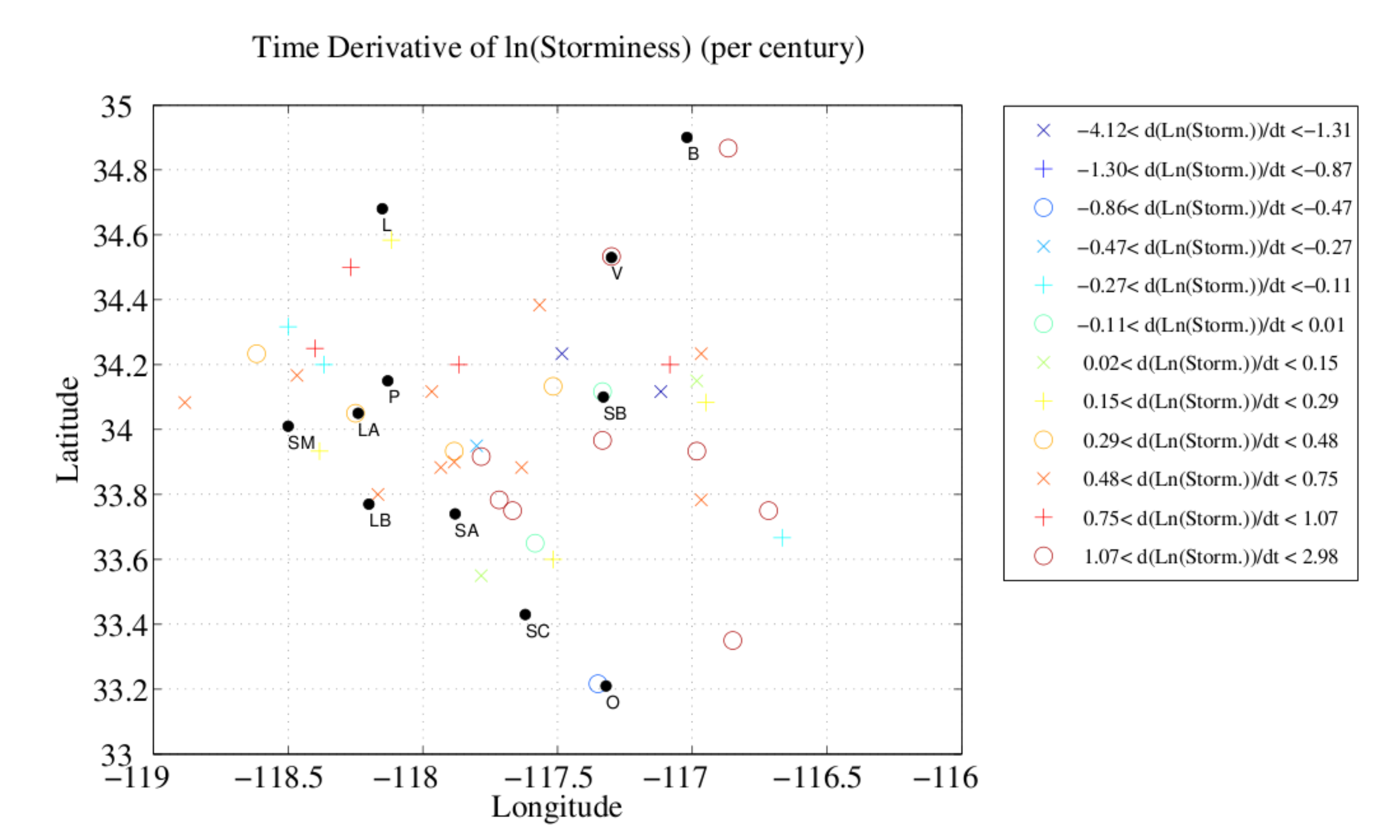}
\caption{\label{LAslopes}Trends in storminess in greater Los Angeles area.
Urban centers are included for reference: B (Barstow), L (Lancaster), V
(Victorville), P (Pomona), SB (San Bernardino), SM (Santa Monica Pier), LA
(Los Angeles City Hall), LB (Long Beach), SA (Santa Ana), SC (San Clemente)
and O (Oceanside); coordinates from Wikipedia.  Symbols have the same
significance as in Fig.~\ref{slopemap}}
\end{figure}

Fig.~\ref{slopemap} shows a striking concentration of positive storminess
slopes in the greater Los Angeles area.  This is shown in more detail in 
Fig.~\ref{LAslopes}.  The boundaries of this region are necessarily
arbitrary, but of the 38 stations inside it, 9 have slopes in the highest
vigintile (5\%) of slopes of the 1722 stations, 13 in the highest decile
and 10 in the second decile.  The {\it a priori\/} probability of 9 in the
highest vigintile is $7 \times 10^{-5}$, of 13 in the highest decile is $4
\times 10^{-5}$ and of 23 in the highest quintile is $5 \times 10^{-8}$.
There are roughly 150 independent regions of similar size in the 48
contiguous states, so the probability of finding such an anomaly
\emph{somewhere} is 150 times as great, but still $\ll 1$\%.  This
geographical concentration of large positive slopes may be
described as no more than a correlation among nearby stations, but the fact
of such a correlation indicates it is a real regional effect, and not an
accidental fluctuation.  It invites speculation about its cause.

\begin{figure}[h!]
\centering
\includegraphics[width=3.6in,angle=-90]{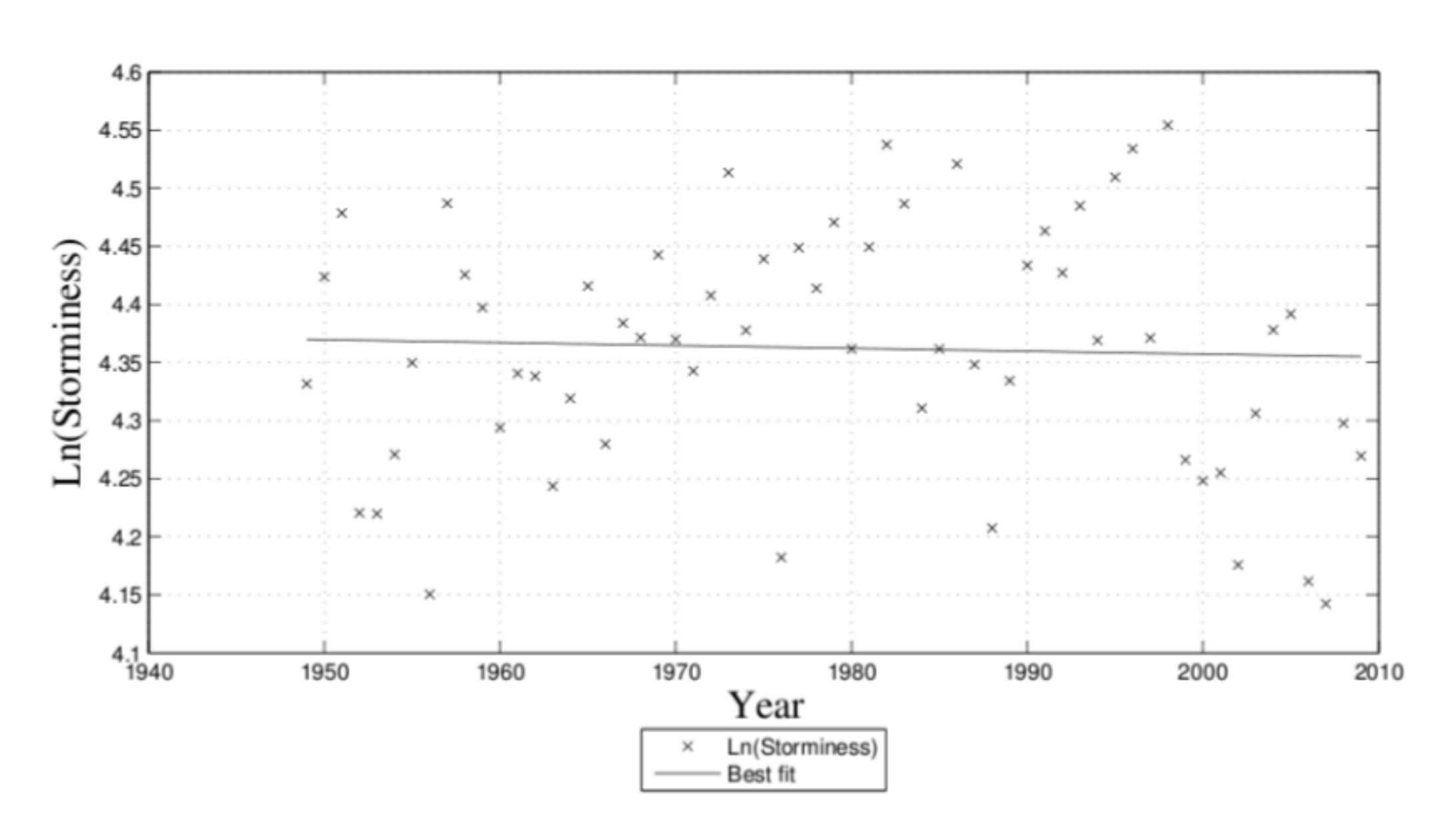}
\caption{\label{byyear}Mean (over stations with sufficient data to be valid)
natural logarithms of storminess {\it vs.\/} year.  No significant trend is
found, setting a lower bound on the characteristic $e$-folding time scale of
$> 500$ years.  This result may be subject to bias if stations with greater
or lesser storminess preferentially have sufficient data to be included in
our analyses.  However, the resulting mean trend is consistent with that
found by first fitting trends to data from each station, and then
averaging.}
\end{figure}

It may also be useful to look for trends in the mean storminess by first
averaging over stations, and then fitting a slope to those averages.  This
inverts the order of analysis shown in 
Figs.~\ref{slopemap}--\ref{scatterplot}, in which the slopes are first
fitted at each station.  In Fig.~\ref{byyear} the mean (over stations with
sufficient data to be valid) natural logarithm of the storminess is computed
for each year.  The logarithms of storminess are used, rather than the
storminesses themselves, so that the mean trend is not dominated by the
stormy stations.  The best linear fit is indicated, with a slope of $(-2.5
\pm 7.7) \times 10^{-4}$/y, consistent with zero.  The nominal $2\sigma$
range of slopes is $(-1.8,1.3) \times 10^{-3}$/y, corresponding to
characteristic $e$-folding times of the storminess of greater than 500
years.  This result is fully consistent with the upper bounds found by first
fitting slopes at each station, as illustrated in
Figs.~\ref{histogram}--\ref{scatterplot}.

\begin{figure}[h!]
\centering
\includegraphics[width=3in,angle=-90]{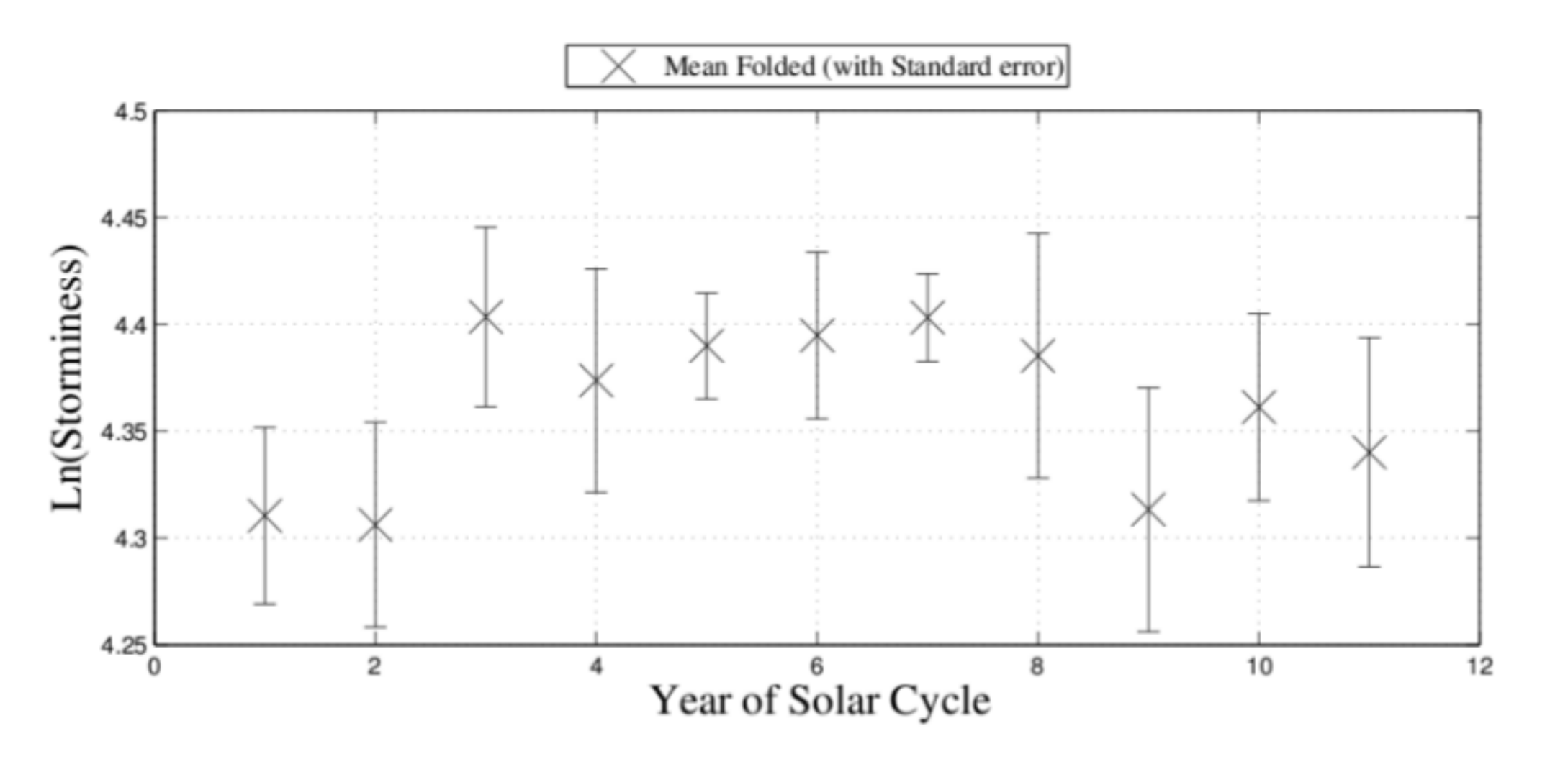}
\caption{\label{solarfold}Mean natural logarithm of storminess, averaged
over stations, folded with an assumed 11.0 year Solar cycle (phase is
arbitrary).  Error bars are $\pm 1\sigma$, based on the distributions of 
averaged (over stations) log-storminesses in the five or six calendar years
corresponding to each year of the Solar cycle.  There is no significant
evidence for an eleven year periodicity.}
\end{figure}

There is a long history of searching for effects of the eleven year Solar
cycle on climate, with controversial results.  We folded the mean (over
stations with sufficient data to be valid) natural logarithms of storminess
shown in Fig.~\ref{byyear} by year of an assumed exact eleven year Solar
cycle (actual times between sunspot maxima or minima have varied by a year
or more from their nominal 11.0 year period).  The results are shown in
Fig.~\ref{solarfold}.  The amplitude of the best fit sine wave is 0.039, but
the null hypothesis of constant log storminess (equal to 4.362, 10 degrees
of freedom) cannot be rejected at the 75\% confidence level ($\chi^2 = 11.4$
with 10 degrees of freedom).

\section{Conclusions}
The data shown here set upper bounds on any long term trend in the mean
storminess, defined as the normalized second moment, averaged over the 48
contiguous United States during the span of our data 1949--2009.  During
that period the mean storminess had a characteristic ($e$-folding) time
scale of variation of not less than about 1000 years at the $2\sigma$ level
of significance, or a corresponding mean rate of change of not more than
0.1\% per year.  There is no evidence for a rapid increase in storminess,
and the mean over stations decreased, though the rate of decrease was not
significantly different from zero.

Extrapolation of these conclusions to the long term future climate depends
on the assumption (necessarily uncertain) that the 60 year period 1949--2009
will be representative of long term secular trends, and was not affected by
shorter term fluctuations analogous to the Little Ice Age.  %  A fit of a
%linear trend (tacitly assuming that the shorter period variations do not
%contribute) sets an upper limit of about 0.002/year to a trend in the
%(dimensionless) scaled normalized second moment, but is consistent with a
%value of zero.
The mean global warming rate over the period 1950--2010 was
$0.010^{\,\circ}$C/yr \citep{GISS11}.   If the natural logarithm of the
normalized second moment of storminess varies in proportion to the warming,
the nominal $2\sigma$ range of their ratio, the sensitivity of
$\ln{(\mathrm{storminess})}$ to global warming, is in the range
$-0.06/^{\circ}$C to $+0.006/^{\circ}$C and
is consistent with zero.  It is not known if this can be extrapolated to
the future, for which climate models predict significant additional warming.

We normalized storminess to the mean precipitation (at each station) over
the entire period 1949--2009, making no allowance for the mean increase of
precipitation of about $7 \times 10^{-4}$/y predicted from the mean warming
rate and the Clausius-Clapeyron equation.  If this assumed mean increase
is applied to the normalizing denominator in Eq.~\ref{norm} (where it is
included to equalize the weighting of trends from stations in different
climate regimes, not to allow for long-term trends in mean precipitation),
the resulting predicted trend in this modified mean storminess would be $-17
\times 10^{-4}$/y, with an uncertainty that depends chiefly on the
uncertainty in the rate of increase of mean precipitation that is difficult
to estimate.  Unlike the storminess defined in Eq.~\ref{norm}, this would
not be a measure of the absolute size of extreme weather events.  

%The rapid rise of the normalized second moment in 2000--2009, a period in
%which the mean temperature rose little, suggests that there is no close or
%causal connection between these measures, at least on decadal time scales.

We can only speculate as to the cause of the comparatively rapid and
statistically significant increase in storminess in the Los Angeles basin
shown in Fig.~\ref{LAslopes}.  During the period covered by our data,
1949--2009, air pollution in that basin was dramatically reduced.
Particulate smog may serve as condensation nuclei for water vapor; if they
are numerous the supersaturated water vapor content of the atmosphere is
divided among many small droplets, producing a gentle drizzle.  Cleaning
the air reduces the number of condensation nuclei, so that the
supersaturated water vapor condenses as fewer but larger droplets, producing
more intense rain and increased storminess.
\bibliographystyle{plainnat}
\bibliography{exweatherLCK}
\end{document}